\documentclass[reprint,superscriptaddress,amsmath,amssymb,aps,prb,floatfix]{revtex4-2}
\usepackage{siunitx}
\usepackage{makecell}
\usepackage[colorlinks,linkcolor=black,anchorcolor=blue, urlcolor=blue,citecolor=blue,unicode]{hyperref}
\usepackage{multirow}
\usepackage{subcaption}
\usepackage[x11names]{xcolor}
\usepackage[normalem]{ulem}
\usepackage{amsmath}
\usepackage{booktabs,siunitx,array}
\usepackage{booktabs}
\usepackage{ulem}
\usepackage{listings}
\usepackage{physics}
\usepackage{import}
\usepackage{xcolor}
\usepackage{framed}
\usepackage{mdwlist}
\usepackage{diagbox}

\usepackage{bm}

\usepackage{graphicx}
\usepackage{dcolumn}

\usepackage{algorithm}
\usepackage{algpseudocode}
\usepackage{csquotes}
\usepackage{array}
\usepackage{siunitx}
\usepackage{amsmath}

\usepackage{ifthen}
\usepackage{threeparttable}
\usepackage{float}

\usepackage{tabularx}
\usepackage{indentfirst}

\newcommand{\ii}{\mathrm{i}}

\begin{document}
\graphicspath{ {article-figure/} }  %

\title{Learning the self-energy through equivariant neural networks
 }
\author{Wudong Shen}
\affiliation{Institute of Physics, Chinese Academy of Sciences, Beijing 100190, China}

\author{Yang Zhong}
\affiliation{Key Laboratory of Computational Physical Sciences (Ministry of Education), Institute of Computational Physical Sciences, State Key Laboratory of Surface Physics, and Department of Physics, Fudan University, Shanghai 200433, China}

\author{Min-Ye Zhang}
\affiliation{Institute of Physics, Chinese Academy of Sciences, Beijing 100190, China}

\author{Xinyang Dong}
\email{dongxy@iphy.ac.cn}
\affiliation{Institute of Physics, Chinese Academy of Sciences, Beijing 100190, China}

\author{Hongjun Xiang}
\email{hxiang@fudan.edu.cn}
\affiliation{Key Laboratory of Computational Physical Sciences (Ministry of Education), Institute of Computational Physical Sciences, State Key Laboratory of Surface Physics, and Department of Physics, Fudan University, Shanghai 200433, China}

\author{Xinguo Ren}
\email{renxg@iphy.ac.cn}
\affiliation{Institute of Physics, Chinese Academy of Sciences, Beijing 100190, China}

\protected\def\xrcolor{\color{red}}

\protected\def\myzcolor{\color{purple}}
\newcommand{\MYZ}[1]{{\myzcolor{\bf #1}}}
\newcommand{\MYZc}[1]{\MYZ{(MYZ: #1)}}
\newcommand{\MYZC}[2]{{\myzcolor\sout{#1}}\MYZ{#2}}
\newcommand{\REV}[1]{#1}
\definecolor{revbcolor}{rgb}{0.00,0.50,0.20}
\newcommand{\REVB}[1]{#1}

\begin{abstract}
The electronic self-energy determines the single-particle Green's function and, through it,
quasiparticle energies and spectral functions, but computing it from first principles is
expensive and it is a difficult machine-learning target: complex-valued, matrix-valued in a
local-orbital basis, and dependent on a continuous frequency. We predict the dynamical
correlation part of the imaginary-frequency \(G_0W_0\) self-energy matrix of molecules from
atomic species and coordinates with an \(E(3)\)-equivariant message-passing network, the
rotational covariance of the orbital-space matrix being imposed by the architecture itself and
the static exchange part taken from the density-functional reference. Frequency enters either
as an explicit network input or through frequency-specific output heads, and the predicted
matrices are continued to the real axis in a representation in which causality and a
non-negative spectral density hold by construction. \REV{The frequency-conditioned model reproduces matrix
elements to below \(1\,\mathrm{meV}\) on the imaginary axis for the systems studied here}, and
that accuracy survives the continuation:
for water clusters, including a tetramer absent from the training set, and for chemically
diverse molecules from QM9, the reconstructed spectral functions follow the \(G_0W_0\)
reference peak by peak and the frontier quasiparticle levels are recovered to a few tens of
meV. Learning the self-energy on the imaginary axis is therefore a practical route to
real-frequency spectra: \REV{once the model is trained, and given the density-functional
reference it is built on, a new geometry costs no explicit \(G_0W_0\) calculation.}

\end{abstract}

\maketitle

\section{Introduction}
\label{sec:introduction}

\REV{First-principles prediction of charged excitations --- quasiparticle energies, band gaps and
spectral functions --- underpins the interpretation of photoemission experiments and the design of
optoelectronic materials.\citep{OnidaG02} Kohn--Sham density-functional theory
(KS-DFT)\citep{HohenbergP64,KohnW65} is the workhorse of electronic-structure theory, but its
eigenvalues are not quasiparticle energies: local and semilocal density-functional approximations
systematically underestimate the band gaps of semiconductors and
insulators,\citep{PerdewJ17GKS} and the formalism gives no direct access to lifetimes or satellite
structures. Generalized Kohn--Sham schemes such as hybrid functionals improve the predicted
gaps\citep{SeidlA96GKS} at the price of adjustable parameters. The \(GW\)
approximation\citep{hedin1965new} addresses the problem at its origin, replacing the static
exchange--correlation potential by the electronic self-energy
\(\Sigma(\mathbf{r},\mathbf{r}',\omega)\) --- complex-valued, frequency-dependent and spatially
nonlocal --- from which quasiparticle energies, lifetimes and spectral functions all
follow,\citep{martin2016interacting,aryasetiawan1998gw} and it describes charged excitations in
molecules and in weakly to moderately correlated materials with broad
success.\citep{reining2018gw,golze2019gw,hybertsen1985first,shishkin2006implementation_clean,jiang2016gw}}

\REV{\(GW\) is by now a standard tool, but it has not become part of high-throughput materials
workflows in the way Kohn--Sham DFT has. \REVB{The reason is cost.} The
self-energy is built from the dynamically screened Coulomb interaction, a nonlocal and
frequency-dependent object that has to be constructed and then convolved with the Green's function
at every point of a frequency grid; even with the low-scaling algorithms developed over the past
decade, a \(GW\) calculation therefore remains considerably more expensive than the ground-state
calculation it starts from.}

\REV{For ground-state and mean-field quantities, machine learning has \REV{substantially alleviated the}
corresponding bottleneck.\citep{schutt2019unifying,gilmer2017neural} Equivariant neural networks
build translational, rotational and permutational symmetries into the architecture, improving data
efficiency, physical consistency and
transferability.\citep{cohen2016group,weiler20183d,thomas2018tensor,brandstetter2021geometric,geiger2022e3nn,batzner20223,battaglia2018relational,unke2021se}
\REVB{Exploiting} the nearsightedness of electronic matter,\citep{prodan2005nearsightedness} the
Kohn--Sham Hamiltonian matrix itself can be learned in a local-orbital representation directly from
the atomic
geometry,\citep{qiao2022informing,su2023efficient,gong2023general,yu2023efficient,yin2024harmonization,yin2024tracegrad,yin2025nextham}
bypassing the self-consistent-field step altogether.\citep{li2022deep} \REVB{A single
\(E(3)\)-equivariant network suffices for transferable orbital-space Hamiltonians in molecules and
solids, hybrid-functional Hamiltonians
included,\citep{zhong2023transferable,tang2024deep_hybrid} and universal models now cover broad
classes of materials, with and without spin--orbit
coupling.\citep{ZhongY24Universal,ZhongY26SOC}}}

\REV{Learning dynamical many-body quantities has accordingly begun. For correlated model systems,
machine learning has been used to represent the Green's function of the Anderson impurity
model,\citep{ArsenaultLF14} to predict impurity spectral functions,\citep{SturmEJ21} and to build a
data-driven dynamical mean-field theory solver.\citep{SheridanE21} For molecules and materials,
machine-learned Green's functions have yielded molecular excitation
spectra,\citep{venturella2023machine_clean} an \(E(3)\)-equivariant network has predicted
self-energy matrices in the Lehmann representation, from which finite-temperature Green's functions
are reconstructed,\citep{dong2024equivariant} and a unified framework predicts the many-body
self-energy at the \(GW\) and coupled-cluster levels.\citep{venturella2025unified} The two closest
to the present work differ from it in ways best stated as trade-offs.
Reference~\citenum{dong2024equivariant} shares the equivariant architecture and the choice of the
self-energy as target, but parametrizes it by a set of poles common to all matrix elements, so that
the analytic structure is imposed by construction at the price of fixing the spectral resolution in
advance; its targets are self-consistent second-order and \(GW\) self-energies on the
finite-temperature Matsubara axis for a small set of species.
\REVB{Reference~\citenum{venturella2025unified} learns the same object in a symmetry-adapted basis
obtained by block-diagonalizing the mean-field Fock matrix, in which the matrix elements are
rotationally invariant; its graph is built on orbitals rather than atoms, and its features are the
converged mean-field matrices themselves. The network therefore emits scalars and needs no
equivariant tensor algebra, which makes it computationally lighter, but it takes a converged
mean-field solution as input rather than the atomic structure alone.}}

\REV{The route taken here is chosen not so much to remedy these differences as to match, on both
sides, the representation in which the calculation is actually carried out. The numeric
atom-centered orbital (NAO) basis is where equivariant message-passing models of the Kohn--Sham
Hamiltonian are most at home: it is compact, localized and sparse, and the orbital-space matrix is
precisely the object such networks are constructed to emit. It is also the basis in which a mature
\textit{ab initio} \(GW\) route operates, in which the self-energy is evaluated on an
imaginary-frequency grid and the real-frequency spectral functions and quasiparticle energies are
obtained from it by analytic
continuation.\citep{ren2012resolution,ren2021all,Zhang2026,gong2026g0w0,jia2026} The grids, the
matrix representation and the continuation step are all established practice there. The question
the present work asks is whether a machine-learned self-energy can reproduce that route end to
end. \REVB{A model that predicts \(\Sigma_{\mathrm c}(\ii\omega_n)\) on the same grid replaces its
single most expensive step while leaving everything downstream --- the Dyson equation, the
continuation and the spectral analysis --- unchanged, and can be trained on, and validated
against, the data these codes already produce.} The imaginary axis is favourable in its own right, \(\Sigma_{\mathrm c}\) being smooth
there and free of the pole structure of the real axis.}

\REV{We accordingly develop an \(E(3)\)-equivariant message-passing framework for the
imaginary-frequency \(G_0W_0\) self-energy of molecules, conditioned only on the atomic species,
the coordinates and \(\omega\). The learning target is the dynamical correlation part
\(\Sigma_{\mathrm c}\), the static exchange part being taken from the density-functional
reference. We target the self-energy rather than the Green's function because \(G_0\) is known
exactly from the preceding DFT calculation, so that the Dyson equation recovers the non-interacting
limit and the analytic structure of \(G\). The target is nonetheless harder than a Kohn--Sham
Hamiltonian: it is complex- and matrix-valued, it depends on a continuous frequency, it is
sensitive to nonlocal screening, and its errors must survive analytic continuation. The reference
calculations reported here are all-electron.}

\REV{We compare two treatments of the frequency dependence. In the fixed-frequency formulation a
shared equivariant backbone feeds one frequency-specific output head for each of the 32 sampled
points; in the frequency-conditioned formulation \(\omega\) is an explicit network input, so that a
single model represents \(\Sigma_{\mathrm c}(\ii\omega)\) as a continuous and differentiable
function that can be queried away from the training grid. The two have parameter counts of the same
order. The predicted matrices are continued to the real axis and inserted into the Dyson equation
to obtain the Green's function, the quasiparticle levels and the spectral function; because the
continuation is ill-conditioned and can turn small matrix-element errors into spurious spectral
features, we perform it in a representation in which causality and a non-negative spectral density
hold by construction, described in Sec.~\ref{sec:analytic_continuation}.}

\REV{The results show that a symmetry-preserving architecture can learn this dynamical,
matrix-valued many-body quantity accurately enough for downstream spectroscopy, and that the
imaginary-axis accuracy survives the continuation. The model is validated on water monomers, dimers
and clusters, including a tetramer absent from the training set, and on the chemically more diverse
QM9 set.\cite{RamakrishnanR14QM9} On QM9 the frequency-conditioned workflow is the more accurate of
the two for the HOMO, the LUMO and the gap, for the representative spectral functions and for both
imaginary-axis matrix diagnostics, the margin ranging from a factor of three on the LUMO to a
factor of eight on the element-wise self-energy error, while the fixed-frequency workflow remains a
useful reference at comparable parameter count. Once either model is trained, obtaining
\(\Sigma_{\mathrm c}\) for a new geometry requires no explicit \(G_0W_0\) calculation.}

The remainder of this paper is organized as follows. Section~\ref{method} introduces the \(G_0W_0\) self-energy on the imaginary-frequency axis, the equivariant network architecture and the two frequency treatments, and the causal analytic-continuation procedure. Section~\ref{sec:computational_details} describes the reference calculations, the datasets and the training protocol. Section~\ref{sec:results} presents benchmark results for water clusters, including an out-of-distribution tetramer, and for molecules drawn from QM9. Section~\ref{sec:summary} summarizes the main conclusions and discusses extensions to periodic systems.

\section{Method}
\label{method}

\begin{figure*}[t]
    \centering
    \includegraphics[width=0.65\textwidth]{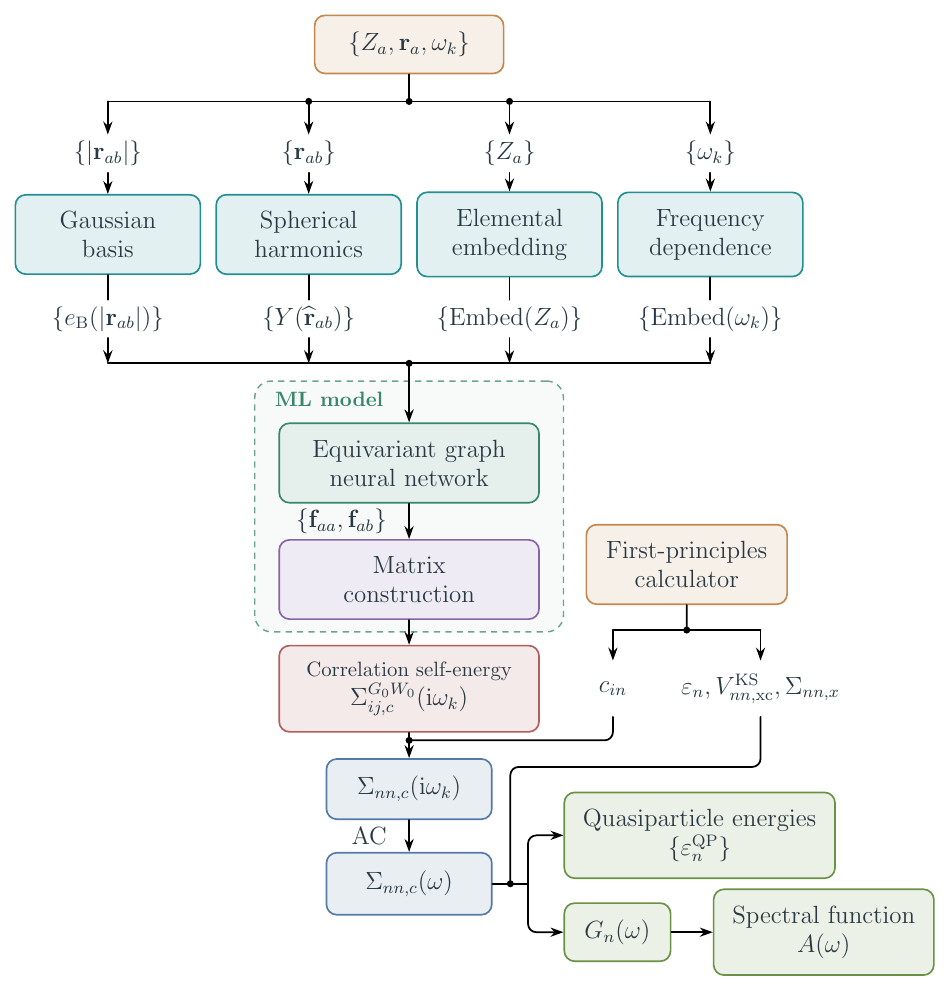}
    \caption{Frequency-conditioned E(3)-equivariant workflow for predicting the correlation self-energy \(\Sigma_{\mathrm c}(\ii\omega_n)\). Atomic species, coordinates and the sampled imaginary frequency are encoded as equivariant features, which the graph neural network and the matrix-construction block map onto the orbital-space self-energy. The prediction is then analytically continued to the real axis and combined with the Kohn--Sham reference to construct \(G(\omega)\), from which the quasiparticle energies and the spectral function follow.}
    \label{fig:nn_architecture}
\end{figure*}

In this section, we present the methodological framework developed in this work. We first review the \(G_0W_0\) formalism for evaluating the electronic self-energy on the imaginary-frequency axis and discuss the properties that make this quantity well suited to machine learning from first-principles data. We then introduce the equivariant message-passing architecture for predicting structure-dependent matrices and describe two strategies for incorporating frequency dependence. Finally, we show how the predicted self-energies are used to reconstruct the Green's function and spectral function, providing a downstream physical validation of the model. \REV{The complete frequency-conditioned workflow is summarized in Fig.~\ref{fig:nn_architecture}.}

\subsection{Dynamical self-energy from the \(G_0W_0\) approach}

In this work we use the widely adopted one-shot \(G_0W_0\) approximation, in which the
noninteracting Green's function \(G_0\) and the screened Coulomb interaction \(W_0\) are both
constructed from a fixed DFT reference. \REV{We work throughout on the imaginary-frequency axis.}
Separating the screened interaction into its bare and induced parts,
\(W_0=v+W_{\mathrm c}\), splits the self-energy into a static exchange term and a dynamical
correlation term,
\begin{equation}
\Sigma^{G_0W_0}(\mathbf r,\mathbf r',\ii\omega)
=
\Sigma_{\mathrm x}(\mathbf r,\mathbf r')
+
\Sigma_{\mathrm c}(\mathbf r,\mathbf r',\ii\omega),
\label{eq:sigma_xc_split}
\end{equation}
where the exchange part is the Fock operator evaluated with the KS orbitals,
\begin{equation}
\Sigma_{\mathrm x}(\mathbf r,\mathbf r')
=
-\sum_{s}^{\mathrm{occ}}
\psi_{s}(\mathbf r)\psi_{s}^{*}(\mathbf r')\,
v(\mathbf r,\mathbf r') ,
\label{eq:sigma_x}
\end{equation}
and the correlation part is the convolution of \(G_0\) with the induced interaction
\(W_{\mathrm c}=W_0-v\),
\begin{equation}
\Sigma_{\mathrm c}(\mathbf r,\mathbf r',\ii\omega)
=
-\frac{1}{2\pi}\int_{-\infty}^{\infty}\!\!d\omega'\,
G_{0}(\mathbf r,\mathbf r',\ii\omega-\ii\omega')\,
W_{\mathrm c}(\mathbf r,\mathbf r',\ii\omega') .
\label{eq:sigma_c}
\end{equation}
Here \(G_0\) is the noninteracting Green's function,
\begin{equation}
G_{0}(\mathbf{r}, \mathbf{r}', \ii\omega)
= \sum_{s} \frac{\psi_{s}(\mathbf{r}) \psi_{s}^{*}(\mathbf{r}')}{\ii\omega + \mu - \epsilon_{s}} \, ,
\label{eq:g0}
\end{equation}
where \(\psi_{s}\) and \(\epsilon_{s}\) are the wave function and orbital energy of the KS
state \(\ket{s}\), and \(\mu\) is the chemical potential, which for the closed-shell molecules
considered here is placed at the centre of the HOMO--LUMO gap.

Only the correlation part \(\Sigma_{\mathrm c}\) is taken as the learning target in this work.
Two of its properties make it a well-conditioned quantity to learn: it decays to zero at large
\(|\omega|\), whereas the full self-energy approaches the constant \(\Sigma_{\mathrm x}\), so
the target has a known asymptotic limit; and its magnitude is not dominated by the large and
essentially structure-independent exchange contribution.

This choice is one of scope rather than of principle. The static part is equally amenable to
machine learning: predicting static mean-field Hamiltonian matrices in a local-orbital
representation is precisely what the Hamiltonian-learning literature
does,\citep{li2022deep,zhong2023transferable,gong2023general} and \(\Sigma_{\mathrm x}\) has the
same matrix structure and obeys the same rotational covariance as those targets. The closest
precedent is the deep-learning treatment of hybrid functionals,\citep{tang2024deep_hybrid}
where the learned Hamiltonian already contains a nonlocal Fock exchange operator of exactly
this form. The complete \(GW\) effective Hamiltonian could therefore in principle be supplied
by two networks, one for the static part and one for \(\Sigma_{\mathrm c}(\ii\omega)\). We do
not pursue that here because the static part is not the computational bottleneck: it follows
from the occupied KS orbitals of the same DFT reference at considerably lower cost, whereas the
dynamical, screening-dependent part is what makes a \(GW\) calculation expensive.
The reference self-energies are evaluated with NAO basis
sets combined with the resolution-of-identity (RI) technique.\cite{ren2012resolution,ren2021all}
Within RI, products of NAOs are expanded in an auxiliary basis,
\begin{equation}
\phi_i(\mathbf r)\phi_j(\mathbf r)
\approx
\sum_{P=1}^{N_{\mathrm{aux}}}
C_{ij}^{P}\,\chi_P(\mathbf r) ,
\label{eq:ri_expansion}
\end{equation}
where \(C_{ij}^{P}\) are the RI expansion coefficients, \(i,j\) label localized NAOs, and
\(P,Q\) label auxiliary basis functions \(\chi_P\). This expansion allows the polarizability
and the screened interaction to be represented in the auxiliary basis. All calculations
reported here use the localized RI variant (RI-LVL),\cite{Ihrig2015,Lin2020LRI} which restricts
the auxiliary functions to the atoms associated with each orbital pair and thereby reduces the
computational cost while retaining high accuracy.
In this representation, the orbital-space correlation self-energy becomes
\begin{equation}
\begin{aligned}
\Sigma_{\mathrm c,ij}(\ii\omega)
=&
-
\int_{-\infty}^{\infty}\frac{d\omega'}{2\pi}
\sum_{sPQ}
\frac{
\widetilde C^P_{is}
W_{\mathrm c,PQ}(\ii\omega')
\widetilde C^{Q*}_{js}
}{
\ii\omega-\ii\omega'+\mu-\epsilon_s
},
\end{aligned}
\label{eq:nao_sigma_matrix}
\end{equation}
where
\(\widetilde C^P_{is}=\sum_j C_{ij}^P c_{js}\)
is the RI coefficient with one index contracted into the KS space through the KS eigenvector
matrix \(c\). The matrices \(\Sigma_{\mathrm c,ij}(\ii\omega_n)\), expressed in the NAO basis,
are the quantities predicted by the neural network; their real and imaginary parts are treated
as two separate real, equivariant targets.

The network therefore supplies \(\Sigma_{\mathrm c}\) and nothing else. Every other ingredient
of the downstream analysis --- the exchange self-energy \(\Sigma_{\mathrm x}\), the KS
eigenvalues and eigenvectors, the overlap matrix, and the exchange-correlation potential
\(V_{\mathrm{xc}}^{\mathrm{KS}}\) --- is taken directly from the same DFT reference from which
\(G_0\) and \(W_0\) are constructed. A ground-state DFT calculation is thus still performed for
each structure; what the model replaces is the \(GW\) step. These static ingredients enter the quasiparticle Hamiltonian
only through the combination \(H^{\mathrm{KS}}-V_{\mathrm{xc}}^{\mathrm{KS}}+\Sigma_{\mathrm x}\),
which is itself an established target for equivariant Hamiltonian learning, so the division is
one of scope rather than of principle; we return to it in Sec.~\ref{sec:summary}.
To make the rotational covariance of these matrices explicit, we resolve the orbital indices as
\(i\equiv(I,p_1,l_1,m_1)\) and \(j\equiv(J,p_2,l_2,m_2)\), where \(I,J\) denote atomic centers, at positions \(\mathbf r_I\),
\(p_1,p_2\) radial channels, and the NAOs are products of a radial function and a real spherical
harmonic,
\(\phi_{Iplm}(\mathbf r)=u_{pl}(|\mathbf r-\mathbf r_I|)\,
Y_{lm}(\widehat{\mathbf r-\mathbf r_I})\).
Let \(\mathcal R\in O(3)\) act on the molecular geometry, so that the atom at
\(\mathbf r_I\) moves to \(\mathcal R\,\mathbf r_I\) while retaining its label, and let
\(\boldsymbol\Sigma'\) denote the correlation self-energy matrix of the rotated structure. Then each orbital-pair block with fixed
\((I,p_1,l_1;J,p_2,l_2)\) satisfies
\begin{equation}
\boldsymbol{\Sigma}^{\prime\,l_1l_2}_{\mathrm c;\,Ip_1,Jp_2}(\ii\omega_n)
=
D^{l_1}(\mathcal R)\,
\boldsymbol{\Sigma}^{l_1l_2}_{\mathrm c;\,Ip_1,Jp_2}(\ii\omega_n)\,
D^{l_2}(\mathcal R)^{\mathsf T} ,
\label{eq:sigma_rotation}
\end{equation}
where \(D^{l}(\mathcal R)\) is the real Wigner rotation matrix acting on the degree-\(l\) real
spherical harmonics, so that \(D^{l}(\mathcal R)^{-1}=D^{l}(\mathcal R)^{\mathsf T}\). The blocks
are in addition invariant under translations of the molecule and under permutations of
identical atoms. Equation~(\ref{eq:sigma_rotation}) is the exact constraint that the
architecture of Sec.~\ref{sec:neurla-network} enforces by construction.
The continuation of the predicted self-energies to the real-frequency axis and the subsequent
reconstruction of spectral quantities are described in Sec.~\ref{sec:analytic_continuation}.

\subsection{Equivariant neural-network architecture and workflow-dependent training}
\label{sec:neurla-network}

\REV{At each imaginary-frequency point \(\ii\omega_n\) the learning target is the orbital-resolved
self-energy matrix \(\Sigma(\ii\omega_n)\), which we predict from the molecular geometry with an
\(E(3)\)-equivariant message-passing network, the middle block of
Fig.~\ref{fig:nn_architecture}. Each molecule is treated as an effectively fully
connected graph whose nodes are the atoms \(I,J\) and whose edges carry the pairwise geometric
relations, the localized orbitals keeping the lowercase indices \(i,j\) of
Sec.~\ref{method}. The equivariant backbone is taken over unchanged from the HamGNN
construction for graph-based prediction of orbital-space
matrices,\citep{zhong2023transferable,schutt2021equivariant,unke2021se} and is summarized below
only as far as the rest of this section requires. What the present target adds to it is of three
kinds. The matrix is complex rather than real, so that its real and imaginary parts are carried by
separate branches throughout. It depends on a continuous frequency argument instead of being
static, which is what the frequency encoding introduced below adds to the node features. And it is
an intermediate quantity rather than an end point: the predicted matrix is passed to the analytic
continuation of Sec.~\ref{sec:analytic_continuation}, so that the accuracy which ultimately matters
is that of the continued spectral quantities and not of the matrix alone.}

\REV{Under a rotation \(\mathcal R\) a feature \(x^{(l)}\) of angular momentum \(l\) transforms with
the Wigner \(D\)-matrix, \(x'_{lm}=\sum_{m'}D^{l}_{mm'}(\mathcal R)x_{lm'}\), and such features are
combined by Clebsch--Gordan tensor products, following the standard tensor-field construction for
three-dimensional equivariant networks,\citep{thomas2018tensor,unke2021se}}
\begin{equation}
\begin{aligned}
z_{\alpha m_3}^{(l_3)}
=&
\sum_{\alpha_1\alpha_2}
W_{\alpha\alpha_1\alpha_2}^{(l_1,l_2,l_3)}
\sum_{m_1m_2}
\langle l_1m_1;l_2m_2|l_3m_3\rangle
\\
&\times
\left[U_1(x_1)\right]_{\alpha_1m_1}^{l_1}
\left[U_2(x_2)\right]_{\alpha_2m_2}^{l_2}.
\end{aligned}
\label{eq:tensor_transform}
\end{equation}
\REV{Here \(\alpha,\alpha_1,\alpha_2\) label feature channels within a given angular-momentum
sector, \(U_1\) and \(U_2\) are trainable channel-mixing maps at fixed \(l_1\) and \(l_2\), and
\(W^{(l_1,l_2,l_3)}\) combines the coupled channels into the output representation. Each atom \(I\)
accordingly carries a set of equivariant features \(\mathbf f_I=\{f^{(l)}_{I\alpha m}\}\) that are
irreducible representations of \(O(3)\), labelled by angular momentum and parity \((l,p)\) with
\(N_{l,p}\) channels each; Eq.~(\ref{eq:tensor_transform}) is the only operation by which they are
combined, and it is retained only when \(\lvert l_1-l_2\rvert\leq l_3\leq l_1+l_2\) and
\(p_3=p_1p_2\).}

\REV{The remaining ingredients are those of Ref.~\citenum{zhong2023transferable} and are stated
here only to fix notation. The features are initialized from a learnable embedding
\(\mathrm{Emb}(Z_I)\) of the atomic species, which populates the even-parity scalar channels
\((l,p)=(0,+1)\) and leaves all others empty; every higher irrep is generated subsequently by
equivariant message passing, \(\mathbf f_I'=\sum_{J}\mathbf f_J\otimes\mathbf e_{IJ}\), in which
\(\otimes\) is the coupling of Eq.~(\ref{eq:tensor_transform}) and the edge features
\(\mathbf e_{IJ}=\Phi(B(r_{IJ}))\,Y(\widehat{\mathbf r}_{IJ})\) combine a learnable radial
embedding of the interatomic distance with the spherical harmonics of the bond direction. Node
features are then updated by residual blocks with a gated nonlinearity, which acts on the scalar
channels and gates the higher-\(l\) channels by a learned scalar, so that the update is nonlinear
without breaking equivariance. On-site and off-site atomic-pair features \(\mathbf f_{II}\) and
\(\mathbf f_{IJ}\) are formed from the updated node features by further tensor-product paths, the
off-site path additionally coupling in the edge feature \(\mathbf e_{IJ}\).}

\REV{The parity label is not redundant, and it is what the present target makes essential. A
spherical harmonic evaluated on an interatomic direction obeys
\(Y_{lm}(-\widehat{\mathbf r})=(-1)^{l}Y_{lm}(\widehat{\mathbf r})\), so the edge features enter
with \(p=(-1)^{l}\); coupling two of them, however, already generates the opposite assignment, the
elementary case being the cross product of two polar vectors, which carries \(l=1\) with
\(p=+1\). Angular momentum alone therefore does not fix the behaviour of a feature under
inversion, and the two labels must be carried separately.}

Finally, an inverse tensor product converts each coupled feature into an orbital-space matrix block
\begin{equation}
M_{IJ}^{l_1l_2}
=
\sum_{l_3=\lvert l_1-l_2\rvert}^{l_1+l_2}
\Pi^{l_1l_2}_{l_3}\!\left[\mathbf f_{IJ}^{(l_3)}\right],
\label{eq:matrix_construction}
\end{equation}
Here \(\mathbf f_{IJ}^{(l_3)}\) is the degree-\(l_3\) part of the pair feature and the operator \(\Pi^{l_1l_2}_{l_3}\) performs a Clebsch--Gordan-based projection from it onto the magnetic components of orbital pairs with angular momenta \(l_1\) and \(l_2\), and is the inverse of the coupling in Eq.~(\ref{eq:tensor_transform}). This construction guarantees the rotational covariance of the resulting matrix blocks and
provides the symmetry-preserving map from molecular geometry to \(\Sigma(\ii\omega_n)\). The
parities are fixed by the same construction rather than chosen: the NAO of angular momentum
\(l\) is a radial function times a real spherical harmonic and therefore carries parity
\((-1)^{l}\), so the block connecting orbitals of angular momenta \(l_1\) and \(l_2\) transforms
with parity \((-1)^{l_1+l_2}\), and every irreducible component extracted from it in
Eq.~(\ref{eq:matrix_construction}) must carry that parity. An \(s\)--\(p\) block thus consists
of a single \((1,-1)\) component and a \(p\)--\(p\) block of \((0,+1)\), \((1,+1)\) and
\((2,+1)\) components. The target therefore dictates which irreducible representations the
output layer may emit, and a network that tracked angular momentum alone could produce blocks
that do not transform as a self-energy matrix at all.
~\citep{zhong2023transferable,unke2021se,dong2024equivariant}

The off-site pair features are directional and are not symmetric under exchange of \(I\) and \(J\), whereas the self-energy matrix is. The blocks \(M_{IJ}\) delivered by Eq.~(\ref{eq:matrix_construction}) are therefore symmetrized,
{
\begin{equation}
\Sigma_{\mathrm c,IJ}(\ii\omega_n)
=
\frac{1}{2}
\left[
M_{IJ}(\ii\omega_n)
+
M_{JI}^{\mathsf T}(\ii\omega_n)
\right].
\label{eq:self_energy_symmetrization}
\end{equation}
}
which for an on-site block reduces to averaging it with its own transpose. The symmetrization is part of the output layer, so \(\Sigma_{\mathrm c,IJ}=\Sigma_{\mathrm c,JI}^{\mathsf T}\) holds for the predicted matrix by construction, and it acts on the real and imaginary parts alike.

\REV{The contrast with the symmetry-adapted, block-diagonal strategy of Ref.~\citenum{venturella2025unified}, outlined in Sec.~\ref{sec:introduction}, is made precise here. There the rotational degrees of freedom are removed from the target by the basis transformation itself, so that the network learns rotation-invariant blocks; here the matrix is predicted directly in the NAO basis and the transformation law of Eq.~(\ref{eq:sigma_rotation}) is imposed at every stage, each operation being a Clebsch--Gordan coupling of irreducible features and the final projection returning blocks that transform with the Wigner \(D\)-matrices of the two orbital angular momenta. Equivariance is thus a property of the architecture rather than of the representation, and applies in the same way to on-site and off-site blocks.}

\REV{The blocks are assembled into the complete orbital-space self-energy matrix used for the Green's-function reconstruction. In the fixed-frequency workflow the shared backbone feeds 32 frequency-specific output heads, one per \(\ii\omega_n\), with separate branches for the real and imaginary parts; the reported model shares all backbone layers, so that the frequency dependence enters only at the final projection.} In the frequency-conditioned workflow the imaginary-frequency coordinate is supplied
to a single network as an additional scalar input. It is encoded on a logarithmic frequency
scale,
{
\begin{equation}
\begin{aligned}
u(\omega)&=u_0+(1-2u_0)\,
\frac{\log\omega-\log\omega_{\min}}{\log\omega_{\max}-\log\omega_{\min}},\\
\left[B_{\omega}(\omega)\right]_k&=\sin\!\left(k\pi\,u(\omega)\right),
\qquad k=1,\ldots,64,
\end{aligned}
\label{eq:frequency_log_basis}
\end{equation}
}
where \(\omega_{\min}\) and \(\omega_{\max}\) are the smallest and the largest sampled
frequencies of the imaginary-frequency grid \REV{specified below in
Eq.~(\ref{eq:imaginary_frequency_grid})}, held fixed rather than recomputed per batch, and the offset \(u_0=0.02\) keeps the mapped coordinate away from the nodes of the sine
basis, at which the feature vector would vanish identically. \REV{The logarithmic coordinate is
dictated by the sampling grid rather than by the self-energy: that grid spans six orders of
magnitude and is approximately
uniform in \(\log\omega\), so \(u(\omega)\) distributes the sampled points evenly over the
interval on which the sine basis varies, and the encoding both separates neighbouring
frequencies and stays smooth between them.} The
resulting vector is mapped by a learnable linear layer onto an even-parity scalar feature,
\begin{equation}
\mathbf f_{\omega}^{(0)}(\omega_n)
=
\operatorname{Linear}
\left[
B_{\omega}(\omega_n)
\right],
\label{eq:frequency_embedding}
\end{equation}
which has the same number of channels as the elemental embedding and is added to it.
\REV{The elemental initialization is thereby replaced by}
\begin{equation}
f^{(0)}_{I\alpha 0}(\omega_n)
=
\left[\mathrm{Emb}(Z_I)\right]_{\alpha}
+
\left[\mathbf f_{\omega}^{(0)}(\omega_n)\right]_{\alpha},
\qquad
\alpha=1,\ldots,N_{0,+1},
\label{eq:frequency_node_input}
\end{equation}
the components with \(l>0\) still vanishing at the input. Since \(\mathbf f_{\omega}^{(0)}\) is an even-parity scalar, this addition leaves the transformation law of the node features untouched, and the frequency dependence is carried into every subsequent operation by the message passing. A single network therefore represents \(\Sigma_{\mathrm c}(\ii\omega)\) as a continuous function of its frequency argument.
 
The two frequency treatments differ not only in how frequency is supplied to the network but also in their training objectives. Both workflows include direct supervision of the self-energy in the NAO basis and an auxiliary loss evaluated after transformation to the KS basis. The frequency-conditioned model additionally includes a frequency-derivative loss that promotes a smooth and accurate dependence on \(\ii\omega\). The two objectives are
\begin{equation}
\begin{aligned}
\mathcal{L}_{\mathrm{fixed}}
&=
\mathcal{L}_{\Sigma}
+
\lambda_{\mathrm{KS}}\mathcal{L}_{\mathrm{KS}},
\\
\mathcal{L}_{\mathrm{cond}}
&=
\mathcal{L}_{\Sigma}
+
\lambda_{\omega}\mathcal{L}_{\omega}
+
\lambda_{\mathrm{KS}}\mathcal{L}_{\mathrm{KS}}.
\end{aligned}
\label{eq:workflow_losses}
\end{equation}
The individual loss terms are defined as
\begin{equation}
\begin{aligned}
\mathcal{L}_{\Sigma}
&=
\operatorname{MAE}
\left(
\Sigma_{\mathrm{NAO}}^{\mathrm{pred}},
\Sigma_{\mathrm{NAO}}^{\mathrm{ref}}
\right),
\\
\mathcal{L}_{\omega}
&=
\operatorname{MAE}
\left(
\frac{\partial\Sigma_{\mathrm{NAO}}^{\mathrm{pred}}}{\partial\omega},
\frac{\partial\Sigma_{\mathrm{NAO}}^{\mathrm{ref}}}{\partial\omega}
\right),
\\
\mathcal{L}_{\mathrm{KS}}
&=
\operatorname{MAE}
\left(
\Sigma_{\mathrm{KS}}^{\mathrm{pred}},
\Sigma_{\mathrm{KS}}^{\mathrm{ref}}
\right).
\end{aligned}
\label{eq:workflow_loss_terms}
\end{equation}
\REV{where \(\operatorname{MAE}(X,Y)=N^{-1}\sum_a|X_a-Y_a|\) runs over all \(N\) matrix
elements, with the real and imaginary parts accumulated separately. All energies are in eV, and
the weights \(\lambda_{\omega}\) and \(\lambda_{\mathrm{KS}}\) are fixed empirically and
absorb the differing units of the three terms.} \(\Sigma_{\mathrm{KS}}\) is obtained by transforming the predicted or reference NAO-basis self-energy into the KS-orbital basis,
\begin{equation}
\Sigma_{\mathrm{KS}}(\ii\omega_n)
=
\mathbf c^{\dagger}\,\Sigma_{\mathrm{NAO}}(\ii\omega_n)\,\mathbf c ,
\label{eq:nao_to_ks}
\end{equation}
\REV{where \(\mathbf c\) holds the KS eigenvectors in the NAO basis, \(\psi_s=\sum_i c_{is}\phi_i\), and satisfies \(\mathbf c^{\dagger}S\mathbf c=\mathbf 1\) with \(S\) the NAO overlap matrix.} The KS-basis loss helps control the propagation of self-energy errors into downstream quasiparticle and spectral quantities and is therefore included in both workflows. By contrast, \(\mathcal{L}_{\omega}\) is used only in the frequency-conditioned workflow, because only this model represents \(\Sigma(\ii\omega)\) as a differentiable function of a continuous frequency input. In the calculations reported below, we set
\(\lambda_{\omega}=5\,\mathrm{eV}\) and \(\lambda_{\mathrm{KS}}=10\).

\subsection{Causal analytic continuation and spectral reconstruction}
\label{sec:analytic_continuation}

Analytic continuation connects the correlation self-energy learned on the imaginary-frequency axis to real-frequency spectral quantities. Throughout this subsection \(\Sigma\) denotes \(\Sigma_{\mathrm c}\); the exchange part is static and enters only through the Dyson equation below.

For self-energies obtained directly from a \(G_0W_0\) calculation this is a well-controlled step. The imaginary-axis data are deterministic and carry no statistical noise, and the rational, or Pad\'e, approximants introduced by Vidberg and Serene\citep{VidbergSerene77} reproduce valence quasiparticle energies robustly; this is established practice in numerical-atom-centered-orbital \(GW\) implementations\citep{ren2012resolution,ren2021all} and underlies the molecular \(G_0W_0\) benchmarks of the GW100 test set.\citep{vanSetten2015GW100,golze2019gw} The Bayesian techniques developed for the stochastic imaginary-time data of quantum Monte Carlo, of which the maximum-entropy method\citep{JarrellGubernatis96MaxEnt} is the best known, address a different problem and are not required here.

What changes in the present context is that the self-energy is predicted rather than computed. Analytic continuation is ill-conditioned with respect to perturbations of the input: the coefficients of a high-order rational approximant depend very sensitively on the sampled values,\citep{Beach2000Pade} so that matrix-element errors small enough to be immaterial on the imaginary axis can be amplified into spurious poles and unphysical structure on the real axis. The prediction error of the network therefore propagates into the spectra in a way that has no counterpart when the reference data are continued directly, and it is this, rather than any deficiency of the Pad\'e construction itself, that motivates the construction introduced below.

Strategies for stabilizing the fit against such perturbations include averaging over approximants of different order,\citep{Schott2016AveragingPade} enforcing causality by construction through a Nevanlinna representation,\citep{Fei2021Nevanlinna} and robust pole estimation.\citep{Huang2023RobustAC} We follow the second of these: rather than fitting a rational function whose poles are free to move, we impose the spectral representation of the self-energy on a fixed grid of real frequencies and solve for non-negative weights, so that causality and a non-negative spectral density hold by construction rather than being recovered by penalization. We also tested the AAA algorithm,\citep{Nakatsukasa2018AAA} the standard modern approach to rational approximation, and found no improvement for the present data.
The continuation is carried out in the Kohn--Sham basis: the predicted matrix is transformed
with Eq.~(\ref{eq:nao_to_ks}) and its diagonal elements \(\Sigma_{\mathrm{c},ss}(\ii\omega_n)\)
are continued independently. Each is represented by its spectral, or Lehmann, form
\begin{equation}
\Sigma_{\mathrm c,ss}(z)
=
\int\!{\rm d}E\,\frac{\rho_s(E)}{z-E},
\qquad
\rho_s(E)\ge0 ,
\label{eq:lehmann_sigma}
\end{equation}
which is discretized on a grid of real frequencies that is fixed in advance and identical for
every matrix element, every structure and both workflows,
\begin{equation}
\Sigma_{\mathrm c,ss}(z)
\simeq
\sum_{j=1}^{N_E}\frac{R_{sj}}{z-E_j},
\qquad
R_{sj}\ge0 .
\label{eq:pole_expansion}
\end{equation}
We use \(N_E=1301\) nodes spanning \(E_j\in[-500,800]\,\mathrm{eV}\) with a uniform spacing
\(\Delta E=1\,\mathrm{eV}\); \REV{this choice is discussed together with the regularization
strength below.}

The residues are determined from the sampled imaginary-frequency values. With the kernel
\(K_{nj}=({\ii\omega_n-E_j})^{-1}\), the real and the imaginary parts of the data are fitted to
the \emph{same} residue vector,
\begin{equation}
\min_{\mathbf R_s\ge0}
\left\|
\begin{bmatrix}w_{\mathrm R}\operatorname{Re}\mathbf K\\[2pt]
w_{\mathrm I}\operatorname{Im}\mathbf K\end{bmatrix}\mathbf R_s
-
\begin{bmatrix}w_{\mathrm R}\operatorname{Re}\boldsymbol\Sigma_s\\[2pt]
w_{\mathrm I}\operatorname{Im}\boldsymbol\Sigma_s\end{bmatrix}
\right\|_2^2
+
\alpha\left\|\mathbf D_2\mathbf R_s\right\|_2^2 ,
\label{eq:nnls_objective}
\end{equation}
with \(w_{\mathrm R}=1\), \(w_{\mathrm I}=1/2\), and \(\mathbf D_2\) the second-difference
operator divided by \(\Delta E^2\), which penalizes rapid variation of the residues from node to
node; we use \REV{\(\alpha=10^{-2}\)}. Equation~(\ref{eq:nnls_objective}) is a non-negative
least-squares problem and is solved as such.

\REV{The spacing \(\Delta E\) and the penalty strength \(\alpha\) are not independent, and were
fixed together by two criteria. The first is independent of the neural network: the reference
\(G_0W_0\) self-energies are continued with a given pair \((\Delta E,\alpha)\) and the resulting
quasiparticle energies are compared with those returned directly by the reference
\(G_0W_0\) implementation of Sec.~\ref{sec:computational_details}, which isolates the
error contributed by the continuation itself. The second compares the continued target with the
continued prediction, and therefore measures how faithfully a given imaginary-axis matrix error is
transmitted to the real axis. Both criteria are optimized by the values used throughout,
\(\Delta E=1\,\mathrm{eV}\) and \(\alpha=10^{-2}\). Coarser spacings leave the quasiparticle
energies and the principal spectral peaks unchanged but degrade the weak satellite features, whose
amplitudes lie an order of magnitude below the quasiparticle peaks, whereas appreciably smaller
penalties admit numerous narrow, unphysical structures in the continued self-energy. The accuracy
floor set by the first criterion is quantified in Sec.~\ref{sec:computational_details}.}

Three properties follow by construction rather than by tuning. First, because the node positions
are fixed, Eq.~(\ref{eq:nnls_objective}) is a convex quadratic program with a unique solution:
there are no local minima, and the instability of rational fits with free denominators --- in
which a denominator zero can approach the real axis --- cannot occur. Second, a single analytic
expression supplies both the real and the imaginary part, so the two satisfy the corresponding
Kramers--Kronig relation automatically. Third, evaluating Eq.~(\ref{eq:pole_expansion}) at
\(z=(\omega-\mu)+\ii\gamma\) with \(\gamma>0\) gives
\begin{equation}
\operatorname{Im}\Sigma^{R}_{\mathrm c,ss}(\omega)
=
-\sum_{j}\frac{R_{sj}\,\gamma}{(\omega-\mu-E_j)^2+\gamma^2}
\ \le\ 0
\label{eq:causality_by_construction}
\end{equation}
for any \(\mathbf R_s\ge0\), so that each continued matrix element is causal and carries
non-negative spectral weight whatever the prediction error. We use \(\gamma=0.5\,\mathrm{eV}\).
Equation~(\ref{eq:causality_by_construction}) is a statement about the continued diagonal
elements; since the off-diagonal elements are not continued, it does not amount to positive
semidefiniteness of the full matrix spectral density. No explicit high-frequency constant is
included in Eq.~(\ref{eq:pole_expansion}).

The number of nodes in Eq.~(\ref{eq:pole_expansion}) is a discretization parameter and not a
measure of model flexibility. The positions \(E_j\) are never adjusted, so refining the grid
approaches the integral in Eq.~(\ref{eq:lehmann_sigma}) rather than adding fitting freedom, and
it is this that keeps the problem convex; a representation with few but freely placed poles
would be neither. What the \(N_\omega\) sampled frequencies determine is correspondingly
limited: each of them probes \(\rho_s\) through a Lorentzian of width \(\omega_n\), the widest
of these exceed the support of \(\rho_s\) and therefore measure essentially the same moment, and
the narrowest are finer than \(\Delta E\). The data consequently fix only a small number of
effective degrees of freedom, with the non-negativity constraint and the smoothness penalty
determining the remainder. We therefore use Eqs.~(\ref{eq:lehmann_sigma}) and
(\ref{eq:pole_expansion}) as a stable and causal representation of the continued self-energy,
and do not interpret the individual residues \(R_{sj}\) as a resolved spectral density.

Quasiparticle energies are obtained from the continued self-energy by solving
\begin{equation}
E^{\mathrm{QP}}_s
=
\varepsilon^{\mathrm{KS}}_s
-
V^{\mathrm{KS}}_{\mathrm{xc},ss}
+
\Sigma_{\mathrm x,ss}
+
\operatorname{Re}\Sigma^{R}_{\mathrm c,ss}\!\left(E^{\mathrm{QP}}_s-\mu\right)
\label{eq:qp_equation}
\end{equation}
by fixed-point iteration. All quantities in what follows are expressed in the orthonormal KS representation, in which only the diagonal of \(\Sigma_{\mathrm c}\) is continued. The resulting self-energy is inserted into the Dyson equation
\begin{equation}
G^{-1}(\omega)
=
G_0^{-1}(\omega)
-
\Sigma_{\mathrm x}
-
\Sigma^{R}_{\mathrm c}(\omega)
+
V_{\mathrm{xc}}^{\mathrm{KS}},
\label{eq:dyson_real_axis}
\end{equation}
where \(V_{\mathrm{xc}}^{\mathrm{KS}}\) removes the exchange-correlation contribution already
included in the DFT reference. The static exchange self-energy \(\Sigma_{\mathrm x}\) of
Eq.~(\ref{eq:sigma_x}) is evaluated directly from the occupied KS orbitals of that reference
rather than predicted by the network, and being frequency independent it requires no analytic
continuation; the continuation is applied to \(\Sigma_{\mathrm c}\) alone. The Green's function is evaluated
slightly off the real axis, at \(\omega+\ii\eta\), with an imaginary shift
\(\eta=0.2\,\mathrm{eV}\); Kohn--Sham states lying more than \(20\,\mathrm{eV}\) above the
chemical potential are omitted from the trace. The two small imaginary parts entering the
calculation play different roles and are not to be confused: \(\gamma\) in
Eq.~(\ref{eq:causality_by_construction}) is the distance of the poles of the continued
self-energy from the real axis, and therefore belongs to \(\Sigma^{R}_{\mathrm c}\) itself,
whereas \(\eta\) broadens the Green's function built from it.
 The predicted NAO-basis matrix is already transpose symmetric by Eq.~(\ref{eq:self_energy_symmetrization}), so no symmetry has to be restored before the transformation to the Kohn--Sham basis and the continuation.
In the orthonormal KS representation, the spectral function is defined as
\begin{equation}
A(\omega)
=
-\frac{1}{\pi}
\operatorname{Im}
\left[
\operatorname{Tr}G(\omega)
\right].
\label{eq:spectral_function_definition}
\end{equation}
Because Eq.~(\ref{eq:spectral_function_definition}) traces over the KS basis, \(A(\omega)\) is a total spectral function, that is, an interacting density of states, rather than a state-resolved quantity. The identical continuation procedure, with identical parameters, is applied to the reference and
to the predicted self-energy of every structure, so that any difference between the resulting
spectra originates in the self-energy itself and not in the continuation. The accuracy of that
procedure can be quantified independently: continuing the reference \(G_0W_0\) self-energies and
comparing the resulting quasiparticle energies with those produced directly by the reference
implementation gives
mean absolute deviations of \(8.5\), \(7.8\) and \(15.5\,\mathrm{meV}\) for the HOMO, the LUMO
and the gap over the QM9 evaluation set. These values are the accuracy floor of the continuation
step, against which the prediction errors of Sec.~\ref{sec:results} should be read.

\section{Computational details}
\label{sec:computational_details}
\subsection{Reference \(G_0W_0\) calculations}
\label{sec:comp_reference}

The reference correlation self-energies were generated with the periodic \(G_0W_0\) implementation in FHI-aims,\REV{\ an all-electron electronic-structure package built on NAOs, in which Hartree--Fock, hybrid-functional and many-body perturbation-theory methods are implemented through the resolution-of-identity technique}.~\cite{FHIaims2026,ren2021all} Each molecule or cluster was placed in a cubic cell with a side length of \(21\,\mathrm{\AA}\),
 and the Brillouin zone was sampled at the \(\Gamma\) point only. The Kohn--Sham starting point was obtained with the PBE exchange-correlation functional,~\cite{PerdewJ96PBE} using scalar-relativistic atomic ZORA and Gaussian smearing of width \(0.010\,\mathrm{eV}\). The self-energy was evaluated on the modified Gauss--Legendre imaginary-frequency
grid used by the FHI-aims \(GW\) implementation,\cite{ren2012resolution}
{
\begin{equation}
\begin{aligned}
\omega_n&=\omega_0\frac{1+x_n}{1-x_n},\\
\omega_0&=0.5\,\mathrm{Ha},\qquad n=1,\ldots,32,
\end{aligned}
\label{eq:imaginary_frequency_grid}
\end{equation}
}
where the \(x_n\) are the nodes of the 32-point Gauss--Legendre quadrature on
\((-1,1)\). The Mo\"bius transformation in Eq.~(\ref{eq:imaginary_frequency_grid}) maps
\((-1,1)\) onto \((0,\infty)\), with the scale parameter \(\omega_0\) fixing the midpoint
\(\omega(x{=}0)=\omega_0\); half of the nodes therefore fall below \(\omega_0=0.5\,\mathrm{Ha}\)
and the grid is much denser at low than at high frequencies. The resulting frequencies span
\(6.85\times10^{-4}\) to \(365\,\mathrm{Ha}\) (\(1.86\times10^{-2}\) to
\(9.93\times10^{3}\,\mathrm{eV}\)). All 32 points are retained and used as training targets
and as inputs to the analytic-continuation procedure. The NAO basis consisted of the minimal basis
of occupied free-atom orbitals augmented by the standard first tier of hydrogen-like
functions, which amounts to \(2s1p\) for H and \(3s2p1d\) for C and O. No higher tiers
were included.
\subsection{Dataset construction and partitioning}
\label{sec:comp_datasets}

FHI-aims delivers \(\Sigma_{\mathrm c}\) in the KS-orbital basis; the NAO-basis learning targets were obtained by inverting Eq.~(\ref{eq:nao_to_ks}), \(\Sigma_{\mathrm{NAO}}=S\mathbf c\,\Sigma_{\mathrm{KS}}\,\mathbf c^{\dagger}S\). The resulting matrices were decomposed into on-site and off-site blocks according to the atomic-orbital indexing, and each structure was represented as an atomistic graph whose nodes are the atoms and whose edges are the atom pairs carrying a nonzero self-energy block; matrix elements with absolute values below \(10^{-8}\,\mathrm{Ha}\) were discarded. Each sample consists of one molecular geometry paired with one imaginary-frequency point, the network inputs being the atomic numbers, the Cartesian coordinates, the pairwise geometric information and \(\omega_n\).
The water geometries were generated by first-principles molecular dynamics with
ABACUS,\REV{\ a density-functional package that combines NAOs with pseudopotentials
for large-scale simulations}.\citep{Li2016ABACUS,Zhou2025ABACUS} Each system was propagated in a fixed cubic cell of
side \(20\,\mathrm{\AA}\) in the canonical ensemble at \(400\,\mathrm{K}\), using a
Nos\'e--Hoover chain with a single thermostat of oscillation frequency
\(0.025\,\mathrm{fs}^{-1}\) and a time step of \(1\,\mathrm{fs}\). Each trajectory ran for
\(5\times10^{4}\) steps (\(50\,\mathrm{ps}\)), of which the first \(5\,\mathrm{ps}\) were
discarded as equilibration. Snapshots were taken at least ten steps apart to reduce temporal
correlation, and near-duplicate geometries were removed by a structural-distance criterion
before the dataset was assembled. The same protocol was used for the monomer, the dimer and the
clusters. The QM9 geometries, by contrast, were taken unchanged from the benchmark
dataset:\citep{RamakrishnanR14QM9} 150 molecules were selected and used with their original
compositions and equilibrium structures, without molecular dynamics or further perturbation.
Reference self-energies for all selected structures were then computed with FHI-aims as
described above.

\REV{Separate datasets were constructed for the water-monomer, water-dimer, water-cluster and QM9
benchmarks. The monomer and dimer datasets contained 200 geometries each, the cluster dataset 100
geometries for each of \(\mathrm{H_2O}\), \((\mathrm{H_2O})_2\) and \((\mathrm{H_2O})_3\),
and the QM9 dataset 150 molecules; with 32 imaginary-frequency points per geometry these amount to
6400, 9600 and 4800 structure--frequency samples, respectively. The tetramer was held out
entirely as an out-of-distribution test system. Unless otherwise stated the \emph{geometries},
rather than the individual frequency samples, were divided at random into training, validation and
test subsets in an 8:1:1 ratio, so that every structure carries all of its frequency points into a
single subset; the QM9 test set is accordingly 15 complete molecules. The real and imaginary parts
of \(\Sigma_{\mathrm c}\) were learned by separate branches throughout, so that the
fixed-frequency workflow uses one shared backbone with 32 output heads per component and the
frequency-conditioned workflow one network per component, with parameter counts of the same
order.}
\subsection{Network architecture, training and post-processing}
\label{sec:comp_training}

The network used three equivariant message-passing layers with feature angular momenta up to \(l_{\max}=6\). The radial basis \(B\) of the edge features consisted of 64 Bessel functions multiplied by a cosine cutoff envelope, and the radial embedding \(\Phi\) was a multilayer perceptron with two hidden layers of 64
neurons each. The atomic graph was built from atomic radii of \(6.0\,\mathrm{Bohr}\), scaled by a factor of
\(1.01\), within an overall interaction cutoff \(r_{\mathrm c}=26\,\mathrm{Bohr}\)
(\(\approx13.8\,\mathrm{\AA}\)); since this exceeds the diameter of every molecule and cluster
studied here, every atom pair carries an
edge, which is the sense in which the graph of Sec.~\ref{sec:neurla-network} is effectively
fully connected. The feature multiplicities \(N_{l,p}\) are not uniform across irreducible
representations. Written in the irreducible-representation notation of
\texttt{e3nn},\citep{geiger2022e3nn} in which \(\texttt{n\textsc{x}}lp\) denotes \(n\) channels of
angular momentum \(l\) and parity \(p\), the hidden representation is
\begin{center}
\footnotesize
\texttt{64x0e+64x0o+32x1o+16x1e+12x2o+25x2e+}\\
\texttt{18x3o+9x3e+4x4o+9x4e+4x5o+4x5e+2x6e,}
\end{center}
amounting to 263 channels and 877 components per atom. Almost half of the channels are scalars,
and since the channel mixing at fixed \((l,p)\) scales as \(N_{l,p}^2\), about three quarters of
those weights sit in the two scalar sectors, while the channels with \(l\ge4\) --- which serve
only as intermediates in the message passing, the largest orbital block requiring \(l\le4\) ---
account for little more than one per cent. Each trained network contains approximately
\(3.9\times10^{7}\) trainable parameters.

The frequency-conditioned model was optimized with \(\mathcal{L}_{\mathrm{cond}}\) and the fixed-frequency reference with \(\mathcal{L}_{\mathrm{fixed}}\). Optimization used AdamW\cite{LoshchilovHutter19AdamW} with AMSGrad, \(\beta_1=0.99\), \(\beta_2=0.999\), \(\epsilon=10^{-8}\) and no weight decay. The learning rate started at \(10^{-2}\) and was halved by a plateau scheduler after five epochs without improvement of the training loss, down to a floor of \(10^{-6}\). Training ran for 200 epochs with an early-stopping patience of 30 epochs, with a batch size of 16 for the QM9 benchmark. Training was performed in single precision on a CPU using 48 threads.
The predicted imaginary-frequency self-energy was analytically continued to the real axis with the procedure of Sec.~\ref{sec:analytic_continuation}, and the continued self-energy was used to construct the Green's function and evaluate the spectral function.

\section{Results}
\label{sec:results}

The dataset construction, first-principles reference calculations, and partitioning protocol are described in Sec.~\ref{sec:computational_details}. Unless stated otherwise, the water-monomer, water-dimer, and water-cluster results below are obtained with the frequency-conditioned workflow, which is the primary model of this work. The fixed-frequency workflow is introduced in the QM9 subsection as a reference for a direct comparison on the more chemically complex dataset.

\subsection{Single water molecule H$_2$O}
\begin{figure}
    \centering
    \includegraphics[width=\linewidth]{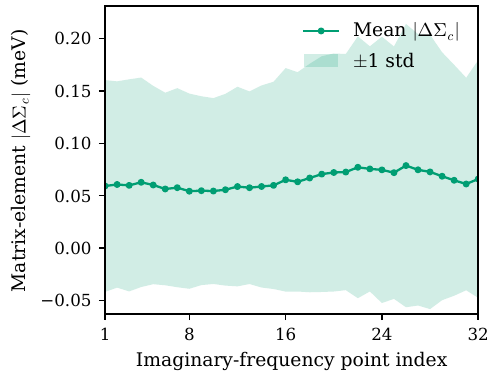}
    \caption{Frequency-resolved H$_2$O correlation self-energy matrix-error statistics
for the frequency-conditioned model. The horizontal axis is the index of the imaginary-frequency
point and the vertical axis the NAO-basis complex matrix-element MAE in meV, with real and
imaginary errors combined as
\(|\Delta\Sigma_{\mathrm c}|=[(\Delta\mathrm{Re}\,\Sigma_{\mathrm c})^2+(\Delta\mathrm{Im}\,\Sigma_{\mathrm c})^2]^{1/2}\).
At each frequency the curve is the mean over the test structures and over all matrix elements,
and the shaded band is \(\pm 1\) standard deviation of the same distribution.
}
    \label{fig:h2o_frequency_error}
\end{figure}

For a single water molecule, we evaluated the frequency-conditioned model on the molecular geometries generated by molecular dynamics (MD) using the ABACUS package.\citep{Li2016ABACUS,Zhou2025ABACUS} The corresponding \(GW\) self-energy matrices were computed with FHI-aims. Figure~\ref{fig:h2o_frequency_error} reports the element-wise absolute error of the complex correlation self-energy matrix in the
NAO basis, with the real and imaginary parts combined as
\(|\Delta\Sigma_{\mathrm c}|=\sqrt{(\Delta\mathrm{Re}\,\Sigma_{\mathrm c})^2+(\Delta\mathrm{Im}\,\Sigma_{\mathrm c})^2}\). At each frequency the average runs over the test structures and over all matrix elements. The mean error stays between about \(0.05\) and \(0.08\,\mathrm{meV}\) across the whole grid,
\REV{with a grid-averaged value of \(0.065\,\mathrm{meV}\),}
and the upper edge of the one-standard-deviation band remains below \(0.22\,\mathrm{meV}\).
Neither grows systematically from one end of the frequency range to the other, although the
spread is somewhat wider at the upper end of the grid.

\begin{figure}[H]
    \centering
    \includegraphics[width=\linewidth]{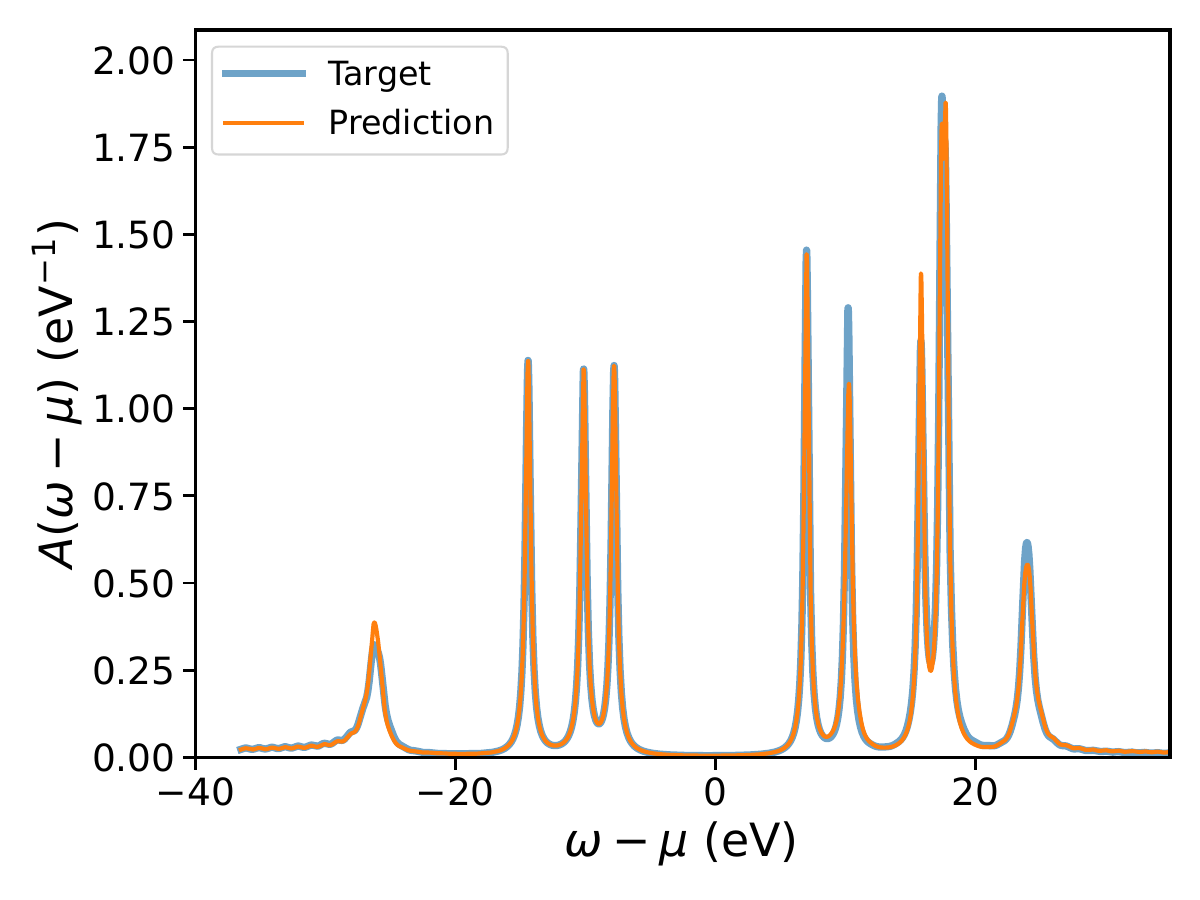}
    \caption{H$_2$O quasiparticle spectral function reconstructed from the target and
predicted correlation self-energies with the causal continuation of
Sec.~\ref{sec:analytic_continuation}. The blue and orange curves denote the \(G_0W_0\) target
and the neural-network prediction, respectively. A spectral broadening of
\(\eta=0.2\,\mathrm{eV}\) is used in the Green's-function
reconstruction.}
    \label{fig:h2o_spectral_comparison}
\end{figure}

As a further validation, we analytically continue the self-energy to the real-frequency axis and construct the Green's function \(G(\omega)\). The spectral function is then evaluated according to Eq.~(\ref{eq:spectral_function_definition}), allowing a direct comparison of quasiparticle spectral functions. \REV{As shown in Fig.~\ref{fig:h2o_spectral_comparison}, the predicted and the reference spectra
are visually indistinguishable across the whole range displayed, in the occupied manifold as well
as above the gap. Each reference peak is reproduced at the same position and with the same weight,
and no spurious structure appears between them: the continuation neither displaces the
quasiparticle poles nor generates satellite features that are absent from the target. This is what
the imaginary-axis errors of Fig.~\ref{fig:h2o_frequency_error} cannot by themselves establish. A
sub-meV matrix-element error is only useful if the continuation does not amplify it, and for this
system it does not.}

The approach therefore works for the simplest system: \(\Sigma_{\mathrm c}\) is reproduced on
the imaginary axis to well below \(1\,\mathrm{meV}\) per matrix element, and --- which is not
automatic, given that the continuation of Sec.~\ref{sec:analytic_continuation} is ill-conditioned
with respect to perturbations of its input --- that accuracy survives the continuation to the
real axis. Two
questions follow. Does the same hold once the orbital space grows and the bonding environment
becomes more varied, and does a model trained on one set of structures carry over to structures
of a kind it has not seen? The next two subsections take these up in turn: first the
water-cluster series, in which the system grows while the chemistry stays the same, and then the
QM9 set, in which the chemical composition itself varies.

\begin{figure}[H]
    \centering
    \includegraphics[width=\linewidth]{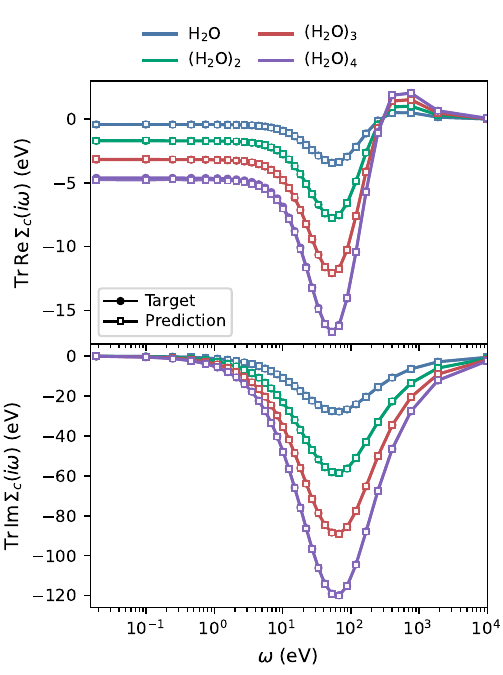}
    \caption{Test-set averaged traces of the correlation self-energy for \(\mathrm{H_2O}\), \((\mathrm{H_2O})_2\), \((\mathrm{H_2O})_3\), and \((\mathrm{H_2O})_4\). For each cluster size, traces from \REV{twenty} randomly selected structures in the corresponding test set are averaged before plotting. The upper and lower panels show \(\mathrm{Tr}\,\mathrm{Re}\,\Sigma_{\mathrm c}(\mathrm{i}\omega)\) and \(\mathrm{Tr}\,\mathrm{Im}\,\Sigma_{\mathrm c}(\mathrm{i}\omega)\), respectively; filled circles and open squares denote the target and prediction. The close agreement of the averaged traces, including for the larger clusters, illustrates the low trace-level error and the model's generalization across molecular geometries and cluster sizes.}
    \label{fig:cluster_trace_comparison}
\end{figure}

\begin{figure}[H]
    \centering
    \includegraphics[width=\linewidth]{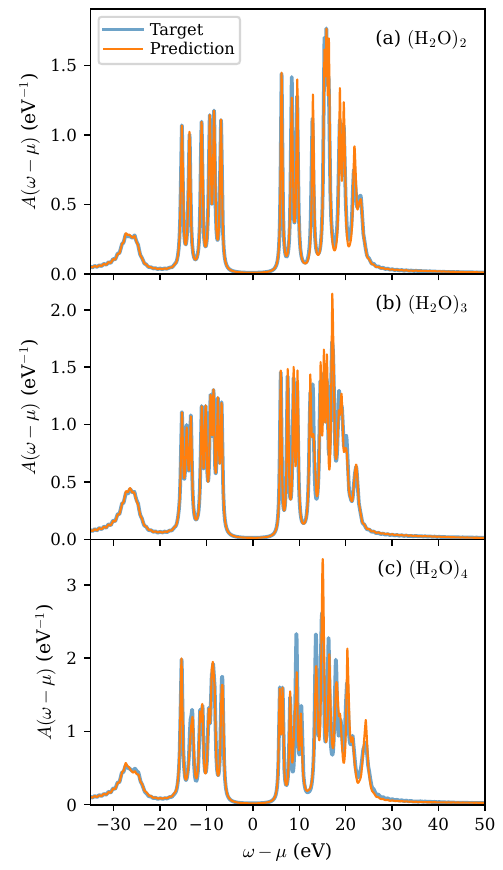}
    \caption{Spectral-function comparisons for the water-cluster systems. Panel (a) shows the water-dimer spectral function corresponding to the dimer result presented in this work, while panels (b) and (c) show the spectra of \((\mathrm{H_2O})_3\) and \((\mathrm{H_2O})_4\), respectively. In each panel, the target and predicted spectra are reconstructed from the \(G_0W_0\) reference and the frequency-conditioned neural-network correlation self-energy. The horizontal coordinate is \(\omega-\mu\). The spectra use the self-energy broadening
\(\gamma=0.5\,\mathrm{eV}\) of Eq.~(\ref{eq:causality_by_construction}) and the Green's-function
imaginary shift \(\eta=0.2\,\mathrm{eV}\).}
    \label{fig:cluster_spectral_comparison}
\end{figure}

\subsection{Water-cluster scaling and out-of-distribution test}
The water-cluster series tests a different question: whether a model trained on the
smaller members of a homologous series transfers to a larger one. Clusters up to the trimer are
represented in the training set, using the datasets and the partitioning protocol of
Sec.~\ref{sec:comp_datasets}. The tetramer is held out entirely, so that its self-energy is
predicted by a model that has never seen a cluster of that size; it probes size transferability
rather than interpolation within a sampled distribution.

Figure~\ref{fig:cluster_trace_comparison} compares the target and predicted traces after
averaging over randomly selected structures from the corresponding test sets. The upper and
lower panels show the real and imaginary trace components, respectively, and both components
follow the reference over the complete imaginary-frequency grid. The mean absolute deviations
between the averaged target and prediction traces are approximately \(0.004\), \(0.021\),
\(0.022\), and \(0.099\,\mathrm{eV}\) for the real part and \(0.004\), \(0.009\), \(0.030\),
and \(0.085\,\mathrm{eV}\) for the imaginary part, for \(\mathrm{H_2O}\), \((\mathrm{H_2O})_2\),
\((\mathrm{H_2O})_3\), and \((\mathrm{H_2O})_4\), respectively.

The trace is extensive, so these values are not comparable across cluster size as they
stand. Per water molecule they become \(0.004\), \(0.011\), \(0.007\) and
\(0.025\,\mathrm{eV}\) for the real part and \(0.004\), \(0.005\), \(0.010\) and
\(0.021\,\mathrm{eV}\) for the imaginary part, so that the per-molecule deviation grows with
cluster size by roughly a factor of five from the monomer to the tetramer. It nevertheless
remains small on the scale of the traces themselves, which reach several eV in the real part and
of order a hundred eV in the imaginary part.
\REV{The trace alone, however, is blind to the off-diagonal elements and admits cancellation between diagonal errors of opposite sign, so we also quote the element-wise error of \(\Sigma_{\mathrm c}\), defined exactly as in Fig.~\ref{fig:h2o_frequency_error}. Averaged over the test structures and over the whole frequency grid it is \(0.065\), \(0.173\), \(0.131\) and \(0.607\,\mathrm{meV}\) for \(\mathrm{H_2O}\) through \((\mathrm{H_2O})_4\). All four lie well below \(1\,\mathrm{meV}\), and the dimer and trimer are within a factor of three of the monomer. The tetramer, which the model has never seen, is about an order of magnitude above the monomer; this is the price of extrapolating beyond the training set, and it is paid on a quantity that is still an order of magnitude smaller than the accuracy floor of the continuation itself.} The accuracy reached on the monomer, dimer and
trimer therefore carries over to a cluster size the model has never seen, with a degradation
that is gradual rather than abrupt --- the behaviour a local, symmetry-preserving representation
of \(\Sigma_{\mathrm c}\) is expected to provide, and what makes the approach useful for systems
larger than those that can be afforded at the \(GW\) level. \REV{One caveat attaches to the trace values: they are deviations of ensemble-averaged traces,
so errors may cancel between structures; the element-wise figures quoted above are free of that
ambiguity.}

The corresponding real-frequency spectra are shown in
Fig.~\ref{fig:cluster_spectral_comparison}. Panel (a) is the water-dimer spectrum, \REV{and panels (b) and (c) those of the trimer and the
tetramer test structures.} \REV{For the dimer and the trimer the predicted and the target curves overlap throughout, as they
did for the monomer. For the tetramer the peak positions remain aligned, but the heights of some
of the stronger peaks differ visibly, the prediction overshooting the reference. The deviation is
therefore one of spectral weight rather than of quasiparticle energy, which is the benign way for
it to appear, since the frontier levels are read from peak positions; and it is consistent with
the element-wise error quoted above being about an order of magnitude larger for the tetramer than
for the monomer.}

\subsection{QM9 Dataset}

We next consider the more complex molecular set drawn from QM9, using the dataset and partitioning protocol described in Sec.~\ref{sec:computational_details}.\cite{RamakrishnanR14QM9}

\begin{figure*}[t]
    \centering
    \includegraphics[width=\textwidth]{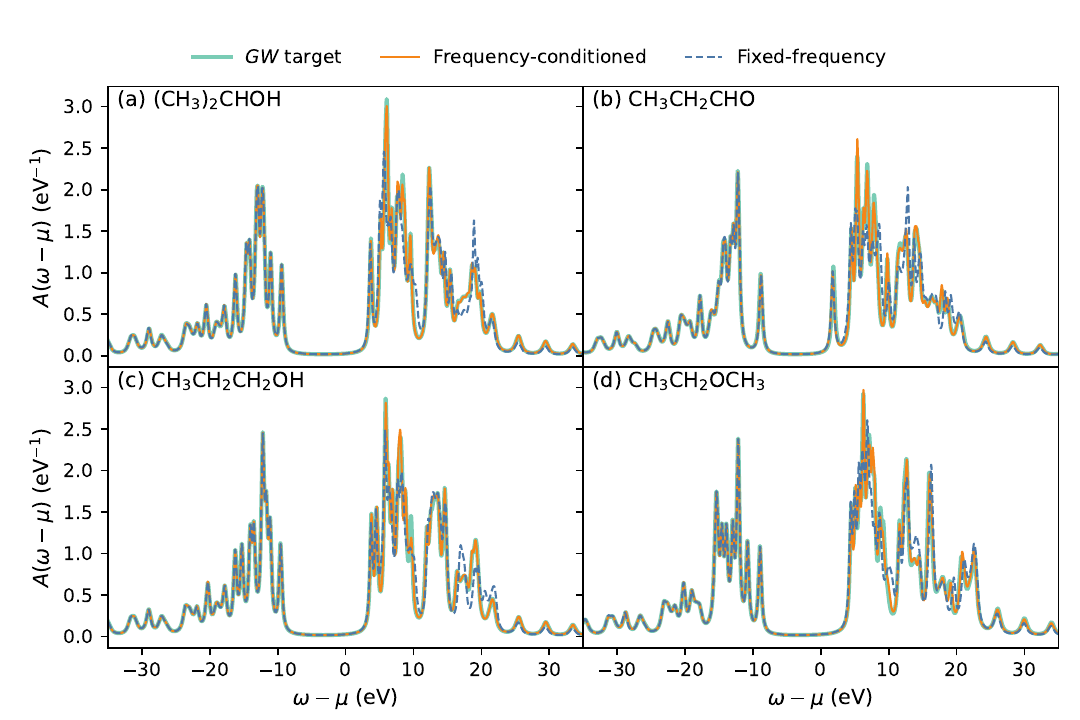}
    \caption{Spectral-function comparisons for four representative QM9 test structures,
obtained with the causal continuation of Sec.~\ref{sec:analytic_continuation}. Panels (a)-(d), in reading order, correspond to 2-propanol, \((\mathrm{CH_3})_2\mathrm{CHOH}\); propanal, \(\mathrm{CH_3CH_2CHO}\); 1-propanol, \(\mathrm{CH_3CH_2CH_2OH}\); and methoxyethane, \(\mathrm{CH_3CH_2OCH_3}\). Each panel compares the \(GW\) target with the predictions of the frequency-conditioned and fixed-frequency workflows.}
    \label{fig:qm9_workflow_spectrum}
\end{figure*}

\begin{figure*}[t]
    \centering
    \includegraphics[width=\textwidth]{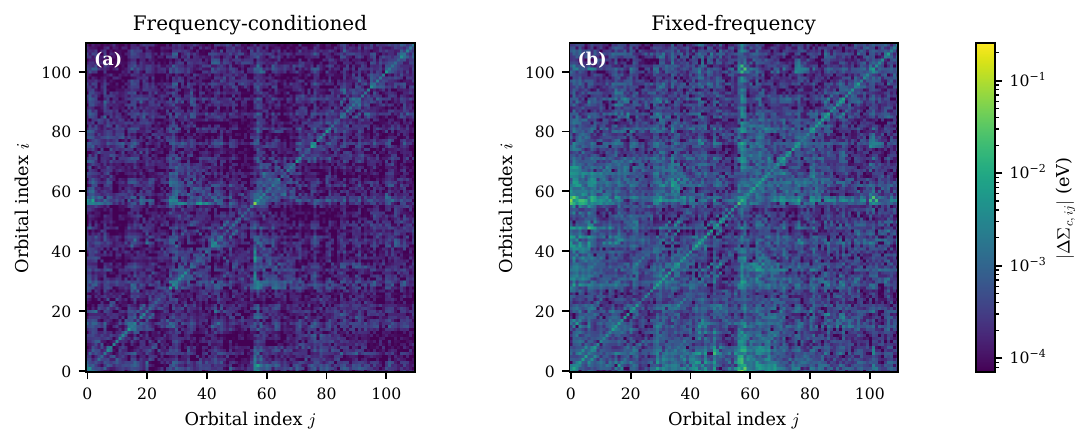}
    \caption{Orbital-resolved absolute correlation self-energy errors for a QM9 \(\mathrm{C}_{4}\mathrm{H}_{8}\mathrm{O}\) test structure at the representative high-frequency point \(\ii\omega_n\), with \(\omega_n=404.43\,\mathrm{eV}\). Panel (a) shows the frequency-conditioned workflow, with a matrix-element MAE of \(0.27\,\mathrm{meV}\), whereas panel (b) shows the fixed-frequency workflow, with a matrix-element MAE of \(0.80\,\mathrm{meV}\). Both panels use the same logarithmic color scale.}
    \label{fig:qm9_error_heatmap}
\end{figure*}

For these larger molecules, which typically contain more than ten atoms, the learning task is substantially more demanding than in the water-cluster tests. The fixed-frequency workflow is compared directly with the frequency-conditioned model here.
\REV{The aggregate statistics and the structure-by-structure comparison below use all fifteen
molecules of the QM9 test set; the full reconstruction pipeline completed on every one of them
for both workflows. Four of the fifteen, drawn at random, are shown as representative spectra in
Fig.~\ref{fig:qm9_workflow_spectrum}, and one as the orbital-resolved map of
Fig.~\ref{fig:qm9_error_heatmap}.}
 Rather than comparing all quasiparticle levels, we focus on the frontier quantities HOMO, LUMO, and the resulting gap, together with representative spectral functions, because these observables are both physically transparent and directly relevant to downstream spectroscopy.

\begin{figure*}[t]
    \centering
    \includegraphics[width=\textwidth]{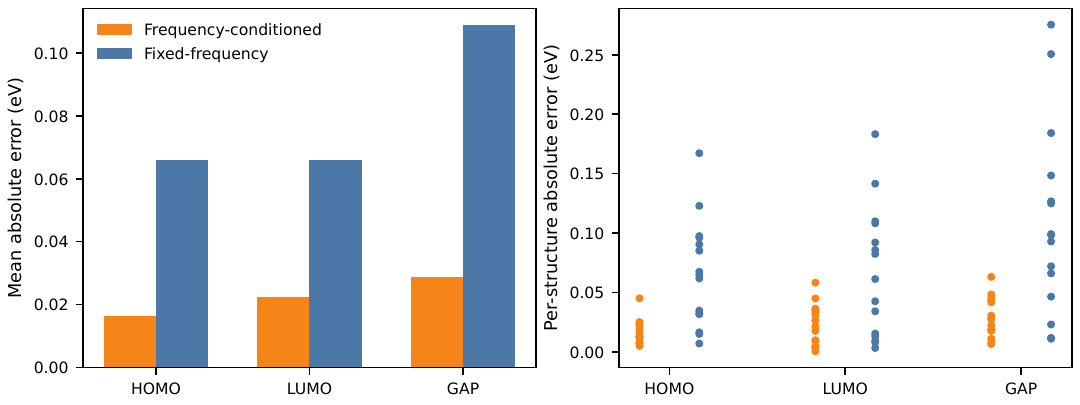}
\caption{Frontier quasiparticle-level errors for the common QM9 evaluation set,
obtained with the causal continuation of Sec.~\ref{sec:analytic_continuation}. The left panel reports the mean absolute errors of HOMO, LUMO, and the gap, while the right panel resolves the same errors structure by structure for the two workflows.}
    \label{fig:qm9_frontier_compare}
\end{figure*}

\REV{Figure~\ref{fig:qm9_frontier_compare} shows that the frequency-conditioned workflow has the
lower mean absolute error for all three frontier observables on the common QM9 evaluation set of
fifteen structures. The mean absolute errors for the frequency-conditioned and fixed-frequency
workflows are \(0.019\) and \(0.065\,\mathrm{eV}\) for the HOMO, \(0.023\) and
\(0.068\,\mathrm{eV}\) for the LUMO, and \(0.027\) and \(0.123\,\mathrm{eV}\) for the gap.
The structure-by-structure panel resolves how uniform this advantage is: the frequency-conditioned
workflow has the smaller error for \(9\), \(10\) and \(12\) of the fifteen test structures for
the HOMO, the LUMO and the gap, respectively, and for \(12\) of fifteen when the three frontier
errors are averaged. The margin is therefore clearest for the gap and least pronounced for the
HOMO, where the two workflows are closer on an appreciable fraction of the structures.}

The same conclusion is visible at the spectral-function level. As shown in Fig.~\ref{fig:qm9_workflow_spectrum}, the frequency-conditioned workflow follows the \(GW\) target more closely across the four structures and reproduces the principal peak positions and spectral weights with smaller deviations. \REV{Across the four panels the curve-wise mean absolute error of the frequency-conditioned
workflow ranges from \(0.009\) to \(0.020\,\mathrm{eV}^{-1}\), against \(0.086\) to
\(0.092\,\mathrm{eV}^{-1}\) for the fixed-frequency workflow; averaged over the four
structures, \(0.016\) versus \(0.089\,\mathrm{eV}^{-1}\), a factor of \(5.6\). The
fixed-frequency error is almost independent of the molecule, whereas that of the
frequency-conditioned model varies with it, which places the residual discrepancy of the
fixed-frequency workflow in its representation of the frequency dependence rather than in the
chemistry of the individual structures.}
To determine whether the distinction between the two workflows is already present at the imaginary-frequency self-energy level, we further compare their matrix-element errors on the same QM9 evaluation set.

\begin{figure*}[t]
    \centering
    \includegraphics[width=\textwidth]{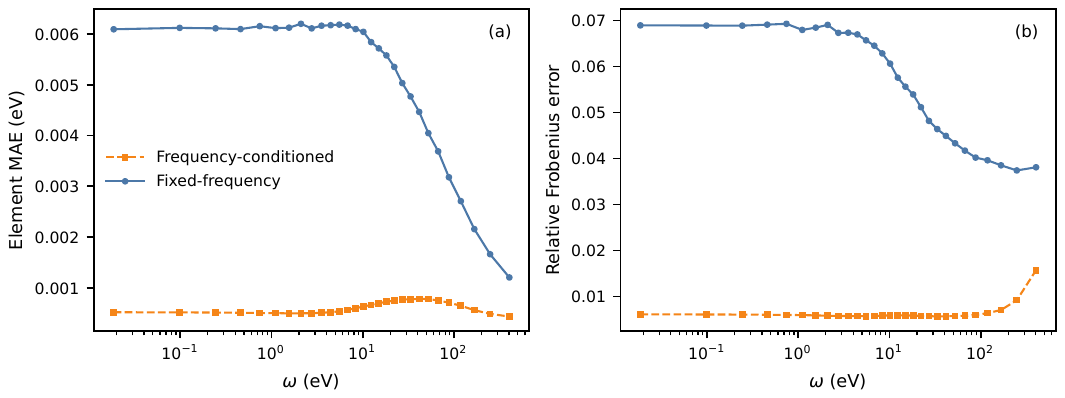}
    \caption{Frequency-resolved comparison of QM9 correlation self-energy errors for the
frequency-conditioned and fixed-frequency workflows on the common evaluation set. Panel (a)
shows the element-wise mean absolute error of \(\Sigma_{\mathrm c}\) in eV, whereas panel (b)
shows its relative Frobenius-norm error.}
    \label{fig:qm9_frequency_error}
\end{figure*}

As shown in Fig.~\ref{fig:qm9_frequency_error}, the frequency-conditioned workflow has the lower error throughout the displayed frequency range in both panels. The mean element-wise MAE is \REV{\(0.55\) versus \(4.55\,\mathrm{meV}\)}, and the mean relative Frobenius error is \(0.0059\) versus \(0.049\), for the frequency-conditioned and fixed-frequency workflows, respectively. Thus, on the reported imaginary-axis diagnostics, the frequency-conditioned model is better by
approximately factors of \(8.3\) and \(8.4\), while also retaining a continuously queryable
frequency representation. The margin is not uniform across the grid. It is widest at low and
intermediate frequencies and narrows at the upper end: in panel (b) the relative Frobenius error
of the frequency-conditioned model rises over the last part of the range, while that of the
fixed-frequency reference falls, so that the two differ by about a factor of two at the highest
sampled frequency rather than by an order of magnitude. This is the region in which
\(\Sigma_{\mathrm c}\) is smallest, and therefore the region that an absolute-error training
objective constrains least; it is also the region that the analytic high-frequency behaviour
discussed in Sec.~\ref{sec:summary} would address.

The orbital-resolved comparison in Fig.~\ref{fig:qm9_error_heatmap} provides a direct view of how these errors are distributed within the self-energy matrix of the \(\mathrm{C}_{4}\mathrm{H}_{8}\mathrm{O}\) test structure. At \(\omega_n=404.43\,\mathrm{eV}\), the matrix-element MAE is \(0.27\,\mathrm{meV}\) for the frequency-conditioned workflow and \(0.80\,\mathrm{meV}\) for the fixed-frequency workflow. Because the two panels share the same logarithmic color scale, the approximately threefold reduction reflects a genuine decrease in the orbital-resolved errors rather than a change in color normalization. This panel is a local diagnostic for one structure and one frequency; the aggregate imaginary-axis comparison is given in Fig.~\ref{fig:qm9_frequency_error}.
\clearpage

The comparisons in Figs.~\ref{fig:qm9_frontier_compare}--\ref{fig:qm9_error_heatmap} consistently favor the frequency-conditioned workflow on the present QM9 evaluation: it has lower mean errors for the HOMO, LUMO, and gap, smaller deviations in the representative spectral functions, and lower values in both reported imaginary-axis matrix-error diagnostics. The orbital-resolved heatmap provides an additional local example, with a matrix-element MAE of \(0.27\,\mathrm{meV}\) versus \(0.80\,\mathrm{meV}\) for the fixed-frequency reference. Thus, explicit frequency conditioning improves the predictive accuracy observed here while also providing a smooth, continuously queryable representation; the shared-backbone fixed-frequency workflow remains a useful reference at comparable parameter scale.

\section{Summary and Prospects}
\label{sec:summary}

\REV{The question posed in Sec.~\ref{sec:introduction} was whether a machine-learned self-energy
can reproduce the route by which \textit{ab initio} \(GW\) is carried out in a numerical
atom-centered orbital basis: evaluate \(\Sigma_{\mathrm c}\) on an imaginary-frequency grid, then
continue it to the real axis. The answer obtained here is that it can. A frequency-conditioned
\(E(3)\)-equivariant network predicts the imaginary-axis correlation self-energy of molecules from
the atomic structure alone to well below \(1\,\mathrm{meV}\) per matrix element, and --- which is
the part that is not automatic --- those matrices survive the continuation: inserted into the same
Dyson equation and the same spectral analysis that the reference calculations use, they reproduce
the \(G_0W_0\) spectral functions peak by peak and the frontier quasiparticle levels to a few tens
of meV, for water clusters including a tetramer absent from the training set and for chemically
diverse molecules from QM9. The learned self-energy is therefore a drop-in replacement for the most
expensive step of an established workflow rather than a parallel construction that has to be
validated on its own terms. Once the model is trained, obtaining \(\Sigma_{\mathrm c}\) for a new
geometry requires no explicit \(G_0W_0\) calculation at all.}

The fixed-frequency formulation is used as a targeted reference rather than as a second main method. Its comparison with the frequency-conditioned model on QM9 shows that explicit frequency conditioning provides both a smooth, continuously queryable representation and lower errors for the frontier levels, representative spectra, and the reported imaginary-axis matrix diagnostics. The shared-backbone fixed-frequency workflow remains a useful reference with a comparable parameter scale, while the present QM9 results support frequency conditioning as the preferred workflow for the downstream observables examined here.

Several directions may further improve the method. The high-frequency behaviour of \(\Sigma_{\mathrm c}(\ii\omega)\) is known analytically,
and imposing the \(1/\omega\) tail as a constraint -- rather than leaving it to be learned from a
grid that is sparsest there -- would target the region in which the present models are least
accurate. A related point concerns the loss: because \(\Sigma_{\mathrm c}\) falls by orders of
magnitude across the grid, an absolute-error criterion is dominated by the low-frequency points,
and a relative error, or a per-frequency normalization of the targets, would give the
high-frequency region a weight commensurate with its role in the analytic continuation. We leave
a systematic study of both to future work. Beyond this, the static and the dynamical parts of the quasiparticle Hamiltonian could be
supplied by two networks trained together. The static ingredients that the present model takes
from the DFT reference enter only through the single combination
\(H^{\mathrm{KS}}-V_{\mathrm{xc}}^{\mathrm{KS}}+\Sigma_{\mathrm x}\), a Hermitian one-body
operator in the same NAO basis and therefore a single equivariant target of exactly the kind
that Hamiltonian models already
predict;\citep{zhong2023transferable,unke2021se} the present model supplies the dynamical part
\(\Sigma_{\mathrm c}(\ii\omega)\). Together with the basis overlap matrix, which is purely
geometric, the two would assemble the full frequency-dependent \(GW\) effective Hamiltonian
\(H^{\mathrm{KS}}-V_{\mathrm{xc}}^{\mathrm{KS}}+\Sigma_{\mathrm x}+\Sigma_{\mathrm c}(\ii\omega)\)
from predicted quantities alone, with no self-consistent-field calculation. A predicted self-energy that is accurate enough on
the imaginary axis further opens the way to self-consistent schemes in which it is re-inserted
into the Dyson equation.
\REV{A further direction concerns the cost of the architecture itself, which is dominated by the Clebsch--Gordan tensor products and grows steeply with \(l_{\max}\). Cheaper equivalents have recently been developed, based either on local rotations or on Gaunt tensor products evaluated with fast Fourier transforms,\citep{PassaroZ23SO2,LiangZ26HotHam} and adapting them to the parity structure of the orbital blocks would reduce the cost of both workflows.}

Extending the present framework to solids is another natural next step. Periodic systems introduce additional complications, such as \(k\)-point sampling and Brillouin-zone integration, which will require corresponding architectural refinements. Reference data for such systems can be generated with low-scaling NAO \(GW\) implementations, which agree with the present all-electron results to within numerical noise while scaling far more favourably for large bulk cells.\cite{Zhang2026}

If successful, adaptation to periodic materials would open the way to high-throughput screening and accurate electronic-structure calculations at a fraction of the cost of conventional ab initio methods. More broadly, it would further integrate many-body physics and symmetry constraints into data-driven electronic-structure modeling.

\begin{acknowledgments}
\REV{We thank Lei Wang and Zechen Tang for helpful discussions.} We acknowledge the funding support from the Advanced Materials--National Science and Technology
Major Project (Grant No.~2025ZD0618401) and the National Natural Science Foundation of China
(Grant Nos.~12374067, 12134012 and 12188101). This work was also supported by the Strategic
Priority Research Program of the Chinese Academy of Sciences under Grant No.~XDB0500201, by the
National Key Research and Development Program of China (Grant Nos.~2022YFA1403800 and
2023YFA1507004), and by the robotic AI-Scientist platform of the Chinese Academy of Sciences.
M.-Y.Z. acknowledges support from the IOP--Humboldt Postdoctoral Fellowship in Physics at the
Institute of Physics, Chinese Academy of Sciences (Grant No.~202402).
\end{acknowledgments}


\bibliography{references}

\end{document}